\documentclass[12pt,onecolumn,twoside]{article}
\usepackage[T1]{fontenc}
\usepackage[utf8]{inputenc}
\usepackage{amsfonts, amsmath}
\usepackage[margin=0.8in]{geometry}
\usepackage{graphicx}
\usepackage{multirow}
\usepackage{lineno}
\usepackage[table]{xcolor}
\usepackage{booktabs}

\usepackage[round,comma,authoryear]{natbib}

\usepackage{authblk}
\author[1,2]{Satavisha De}
\author[1,3,*]{B. Vibishan}
\affil[1]{Department of Biology, Indian Institute of Science Education and Research (IISER), Pune, India}
\affil[2]{Present address: Department of Integrative Biology, University of Texas at Austin, US}
\affil[3]{Present address: Department of Bioengineering, Indian Institute of Science (IISc), Bengaluru, India}
\affil[*]{Corresponding author: vibishanb@iisc.ac.in}

\title{A mechanistic interpretation of patterns in mutation risk across biological scales}
\date{\empty}

\usepackage{hyperref}
\hypersetup{
	colorlinks = true, 
	urlcolor = blue, 
	linkcolor = red, 
	citecolor = red 
}

\begin{document}

\maketitle

Abstract word count: 258 words\\

Text word count: 7653\\

Number of figures: 6\\

Number of tables: 2

\newpage
\section*{\centering Abstract}
\textbf{Background and objectives}: The allometry of cancer risk in mammals is scale-specific, as the inter-specific, intra-specific, and inter-organ patterns in cancer risk scaling are distinct from each other. While cancer risk barely changes with increasing body sizes across species, larger conspecifics are known to be at a higher risk of cancer. Cancer risk trends with age and lifespan have likewise been documented in humans and other species. Nevertheless, mammals broadly share developmental features, and the twin processes of cell division and differentiation are central to developmental dynamics. In this study, we suppose that a better understanding of these processes across biological scales could lead to a clearer understanding of cancer risk allometries. 

\textbf{Methodology}: We develop an agent-based model in which a cell compartment grows from one cell to a target size. We assess how developmental dynamics scale with compartment size and influence mutation accumulation. 

\textbf{Results}: Our data identify specific regions of the parameter space where the accumulation of neutral mutations increases with cell number, while providing a mechanistic explanation for such behavior. The accumulation of non-neutral mutations reveals an interesting antagonism in cell fitness between mutation and differentiation, which could change the viability of a given developmental strategy. Our model is able to produce a wide range of trends in cancer risk allometries in different parts of its parameter space, reflecting the observed heterogeneity across biological scales. 

\textbf{Implications}: Our results provide a theoretical framework that allows us to elucidate the conditions of mutation-differentiation balance under which having more cells does lead to higher cancer risk, and importantly, the conditions under which it does not.

\paragraph{Keywords:} Cancer evolution; Peto's paradox; Allometry; Evolutionary medicine; Agent-based modelling

\section*{\centering Lay summary}
Cancer is a disease of uncontrolled cell division, but whether having more cells increases cancer risk remains an open question. Here, we identify potential roles for developmental mechanisms in explaining how and why cancer risk changes with the number of cells available to become cancerous.

\section{Introduction}
\paragraph{\empty}Mutations are widely considered a basic substrate for cancer growth \citep{Stratton2009a, Nik-Zainal2012, gerlinger2014, Martincorena2015a, Ma2018}. Given that mutation is intrinsically linked to cell division, a question of fundamental importance in cancer biology is whether or not higher cell number and/or more cell division leads to an increase in the risk of cancer. The question arises essentially from the expectation that more cells and/or more frequent cell division present greater opportunities for mutations to occur that could lead to cancerous growth \citep{Peto1975, peto_quantitative_2015}. Given that cell number varies over several orders of magnitude in biology, the scaling of cancer risk (as reflected by mutation risk) can be studied over at least four spatio-temporal scales of biological organization. Across these scales, trends in cancer risk appear to be remarkably divergent. Across species, larger organisms with potentially higher cell number are not at a proportionately higher risk of cancer, nor are longer-lived organisms with potentially more lifetime cell divisions. In fact, the correlation between species cancer risk and body size/lifespan is very nearly nil \citep{Abegglen2015, Gorbunova2014}. Within a given species, however, cancer risk does seem to scale with body size, as larger individuals among both humans and domesticated canines show a higher lifetime incidence of cancer \citep{albanes_adult_1988, green2011, nunney2018, rafalko2023}. At the sub-organismal level in humans, different tissues and organs show variation in cancer risk that can be associated with their cell number to some extent \citep{tomasetti2015, Tomasetti2017}, although exposure to environmental risk factors is thought to play an important role in determining the eventual likelihood of carcinogenesis \citep{Hochberg2017}. Finally, trends of cancer incidence with age in the human population have been the subject of a rich pedigree of epidemiological inquiry \citep{Frank2007}, which has demonstrated that cancer risk saturates at very old age in humans. This indicates that the effect of longer lifespan and a consequent increase in lifetime cell divisions on cancer risk may not be straightforward \citep[see][\empty]{vibishan_context-dependent_2020}. There is therefore considerable, unexplained diversity in cancer risk trends across scales of biological organization. In the current study, we suggest that this diversity of trends could be addressed comprehensively by investigating common processes that are known to act across these scales, which could hold the key to explaining how cancer risk trends vary across scales. Identifying such common processes begins with the recognition that cancer is essentially a competitive phenomenon between cells of varying fitnesses \citep{madan_cell_2022}, and a vast diversity of cellular processes may be involved in somatic evolution of mutant clones that can lead to cancer \citep{gerlinger2014}. This is true of cancerous growth in any organism, across the scales mentioned above, and therefore, processes that affect cell fitness and the dynamics of somatic evolution are generally applicable across those scales. As mentioned earlier, mutations are an important substrate for cancer growth, and this stems from their effect on somatic cell fitness. The other process that could be considered a cell fitness determinant in the soma is differentiation.

\paragraph{\empty}Differentiation is the establishment of a hierarchical tissue architecture that separates cells into stem and differentiated cell pools, with a number of transit stages in between the two. This creates a gradient of division rates, such that stem cells can divide indefinitely, while transit and differentiated cells are typically limited in the number of division cycles they undergo \citep{pepperAnimalCellDifferentiation2007}. Mutation and differentiation can also interact in a variety of ways. This includes mutations that can attenuate the rate of differentiation or induce de-differentiation, thereby enabling unrestricted cell growth \citep{Varga}. On the other hand, mutations that occur in a differentiation stage with a very limited division potential could restrict the extent to which the mutant clone can expand in the cell pool \citep{demeter2022}. Both mutation and differentiation can therefore affect the relative frequencies of mutant and non-mutant cells in a given cell pool, and the overall risk of mutation accumulation and cancer progression is a composite function of both processes. In the context of cancer prevention, empirical evidence suggests that reducing overall mutation risk or increasing the efficiency of DNA repair can both be effective at reducing cancer risk \citep{Tian2013, Sulak2016, vazquezPervasiveDuplicationTumor2021}. On the other hand, theoretical models have explored the possibility that the differentiation hierarchy could enable cancer prevention by restricting the spread of mutant clones \citep{werner2011, Werner2013a, hindersin_should_2016, derenyi2017, derenyi_cancer_2020, grajzel2020}, although direct empirical evidence remains lacking.

\paragraph{\empty}One such theoretical study by Erten and Kokko \citep[\empty][ hereafter referred to as EK20]{Erten2020}, has used an individual-based model of metazoan ontogeny to arrive at optimum differentiation/division strategies that can support viable development of a generic cell compartment starting from a single zygote with a pre-determined target size in terms of cell number at maturity. The EK20 model predicts that high target sizes, i.e., larger organisms,, would be expected to evolve lower rates of differentiation and more transit stages from stem to differentiated cells, indicating that larger stem cell compartments could be a viable strategy to achieve and maintain a higher target size. EK20 also highlight that in metazoan development, an organism of a given size may be viably produced by a variety of parameter combinations due to the multiplicity of parameters involved in development. In turn, these parameter combinations could produce a range of cellular compositions for a given body size with varying levels of robustness to mutation accumulation and cancer risk.

\paragraph{\empty}On the one hand, EK20 has led to several interesting predictions that inform optimal ontogenetic strategies for metazoan development over a range of target sizes across species. On the other hand, the known cancer risk trends across various scales suggests that the optimality of ontogenetic strategies inferred at the interspecific level may not be generalizable to scaling of mutation risk with target size across other biological scales. Furthermore, while EK20 account for mutational damage-induced cell death, the framework does not include a biologically-relevant distribution of fitness effects (DFE) of mutations occurring over the course of ontogeny. Mutations in a somatic context fall on a spectrum of fitness, from deleterious to beneficial in the context of cell growth and cancer \citep{loewe2006, eyre-walker2007}. A DFE that can reproduce such a spectrum of mutations \textit{in silico} could thus lead to a holistic understanding of ontogeny and cancer risk. This may be particularly relevant to EK20's prediction that evolutionary increases in target size could be supported by a concurrent increase in the proportion of undifferentiated stem cells, as a larger stem cell pool also has higher overall self-renewal capacity and could therefore also be at a higher risk of mutation accumulation. This highlights the potential for reciprocal interactions between mutation and differentiation, and it is unclear how the viability of developmental strategies evolved in the EK20 would be affected by the accumulation of non-neutral mutations for a range of target sizes. 

\paragraph{\empty}The current study is an exploration of this balance between neutral and non-neutral mutation and differentiation through two related questions: (1) how do mutation and differentiation affect development and viability in a generic cell compartment of a given size?, and (2) how can the scaling of mutation risk with different target sizes be understood in terms of the balance between differentiation and mutation? To this end, we construct an individual-based model (IBM) of the development of a cellular compartment from a single cell up to a specified target size, with cell division modeled as a function of both differentiation and mutation-driven fitness effects. While our framework is based on that of EK20, we use it to develop a mechanistic understanding of the interplay between dynamics of mutation, cell division, and differentiation without restricting the results to any one level of biological organization. Our results demonstrate how the extent of cell differentiation affects both the size of the actively-dividing cell pool and the average division frequency required to produce a mature cell compartment of a given target size. We also relate this to trends in neutral mutation accumulation for a range of differentiation rates and target sizes, which is the first systematic theoretical report of its kind on neutral mutation accumulation. Distinct parameter combinations in the model can be classified as ``high'' vs ``low'' differentiation ontogenetic strategies, and these differ markedly in their propensity for neutral mutation accumulation, with ``high'' differentiation strategies leading to higher mutation risk. We then show that the viability of a differentiation strategy across target sizes can change due to non-neutral mutation accumulation, as mutation-induced increase in cell fitness can sometimes ``rescue'' certain clonal lineages to ensure viable development. These results are finally discussed from the level of organs to species, to ask which model parameters are necessary, sufficient, or insufficient to explain the cancer risk allometry at each level. Taken together, our results provide a theoretical framework that allows us to elucidate, across levels of biological organization, the conditions under which greater cell number does lead to higher cancer risk through mutation, and importantly, the conditions under which it does not. 

\section{Methods}

\paragraph{\empty}We discuss the implementation of the model and its main findings in terms of a generic cell compartment, and then discuss the implications of our findings at each specific biological level in a later section.

\paragraph{\empty}Figure \ref{fig1}A shows a simplified scheme of the sequence of events in the simulation, and the other subpanels illustrate the main processes and features of the model that govern the growth and composition of the cell compartment. Each individual cell is described by three attributes-fitness, differentiation stage and number of mutation. These attributes are affected by model processes including division, differentiation and mutation. We now detail the parametrization of each of these processes following the overall sequence of events in Figure \ref{fig1}A.

\subsection{Cell fitness}
\paragraph{\empty}We begin the simulation with a generic cell compartment containing a single non-mutant cell in the least differentiated (stem) stage. This cell, upon division, or division with differentiation, can give rise to stem or differentiated cells respectively. We define $\omega$ as the cell fitness, such that cells with higher $\omega$ are more likely to divide. Since stem and non-stem cells differ in their division rates \citep{Varga}, we assume $\omega_1 = 1$, corresponding to the fitness of a stem cell in the least differentiated stage. A given stem cell must go through $T-1$ transit stages to reach the terminally-differentiated state and the transition of a cell from one transit stage to the next leads to a change in its fitness. Accordingly, we take $\omega_i$ to be the cell fitness in differentiation stage $i$, which varies with the differentiation state as follows:
\begin{equation}
	\label{eq1}
	\omega_i = \alpha \times \omega_{i-1}\ \forall\ 1<i \leq T-1,
\end{equation}

Equation \ref{eq1} shows that the effect of differentiation on cell fitness is mediated by $\alpha$. We define $\alpha$ as the gradient of cell fitness with differentiation and use this gradient to describe the full range of theoretical possibilities. Cell fitness can increase, decrease or remain unchanged with differentiation, and it is clear from Equation \ref{eq1} that this corresponds to $\alpha>1$, $\alpha<1$, and $\alpha=1$ respectively (Figure \ref{fig1}B). 

\subsection{Cell division}
\paragraph{\empty}We use the cell fitness $\omega_i$ as the basis of division in the model. We define the divisible cell pool as $S(t) = \sum_{i}^{T-1} n_i$, where $n_i$ is the number of cells in differentiation stage $i$. Cells in stage $T$ are considered terminally-differentiated and not included in the divisible pool. In each iteration of the model, the target size of the compartment, $M$, is used to calculate the required number of cells to fill the compartment, hereafter referred to as the deficit in cell number, $D(t)$. We define $D$ as the scaled difference between the target compartment size and the current number of terminally-differentiated cells in the organism, $0.05 \times [M-n_T(t)]$, where $n_T$ is the number of cells in the $T$ stage. The $0.05$ scaling factor is included following EK20 to produce approximately sigmoid organismal growth \citep{karkach2006}, while including only $n_T(t)$ in the deficit calculation reflects growth dynamics as dependent on fully differentiated, ``functional" cells. If the deficit $D(t)$ is more than the total number of divisible cells $S(t)$, then all $S$ cells undergo division regardless of their fitness. If $D(t)<S(t)$, then a Bernoulli trial is carried out for each cell in the simulation with the probability of success defined as the relative fitness of the focal cell over the most fit cell in the pool, $\frac{\omega_{ij}}{\text{max}(\omega)}$ for the $j$th cell in differentiation stage $i$. Among those cells that succeed in this trial, the first $D$ cells with the highest fitness are chosen to undergo division. 

\subsection{Cell differentiation}
\begin{figure}[ht]
	\centering
	\includegraphics[width = \columnwidth]{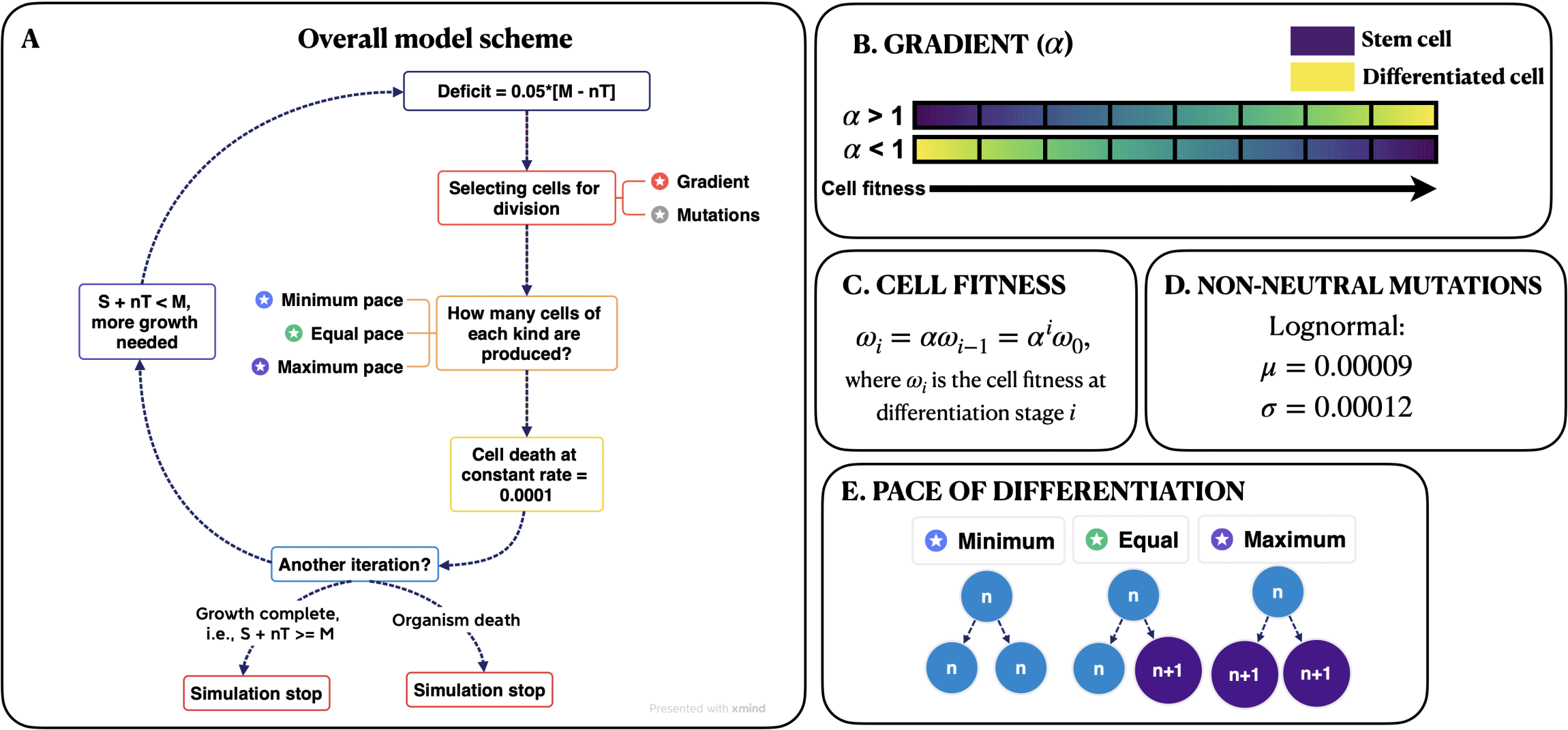}
	\caption[Sequence of events in the model and model components]{\label{fig1}\textbf{Sequence of events in the model and model components}. (A) A simplified scheme of the sequence of events in the model. (B) the effect of gradient $\alpha$ on cell fitness-for $\alpha < 1$, selection favors stem cells while for $\alpha > 1$, selection favors differentiated cells. (C) Exponential scaling of cell fitness with differentiation. (D) Fitness effects for non-neutral mutations are sampled from a lognormal distribution of given mean and SD. (E) The pace of differentiation is simulated at three levels; specific parameter values and the corresponding average number of differentiated cells produced per cell division are given in Table \ref{table1}. All notations are explained in the main text.}
\end{figure}

\paragraph{\empty}Among the cells selected for division, division can occur in one of three ways with respect to the differentiation stages of the progeny cells (Figure \ref{fig1}E). When division is asymmetric, one progeny cell differentiates to the next stage while the other remains in the same stage as the parent. When division is symmetric, both progeny cells are in the same differentiation stage, and both could either differentiate to the next stage or stay in the same stage as the parent. Two probability terms are used to implement these three possibilities stochastically. $P$ is defined as the probability of asymmetric cell division and $Q$ as the probability of no differentiation given that the division is symmetric. This implies that the three kinds of divisions outlined above can be expressed mathematically as: $P$, corresponding to asymmetric division, $(1-P) \times Q$, corresponding to symmetric division where both progeny cells remain in the same differentiation stage as the parent, and $(1-P) \times (1-Q)$, corresponding to symmetric division with both progeny cells differentiating into the next stage. Since the values of $P$ and $Q$ together determine how fast differentiation proceeds with cell division, we considered three different paces of differentiation in terms of combinations of $P$ and $Q$ values. We call these three levels of pace of differentiation-maximum, equal, and minimum-corresponding to whether cell division on average accelerates, retards, or does not affect differentiation, respectively. Table \ref{table1} demonstrates the effective difference between the levels of pace with a numerical test case based on how many progeny cells are produced in differentiation stages $i$ and $i+1$ for every 100 parent cells dividing in differentiation stage $i$. As with cell division, we use the probabilities $P$ and $Q$ in sequential Bernoulli trials for each of the cells selected for division, first to decide if the cell divides asymmetrically, and if not, whether or not differentiation occurs. We then update the numbers of cells in the corresponding differentiation stages based on the outcomes of these trials, and adjust cell fitness according to Equation \ref{eq1}.

\subsection{Mutation}
\paragraph{\empty}The primordial stem cell at the beginning of the simulation starts with no mutations. After that, mutations occur probabilistically with every cell division event. We use a fixed mutation rate per cell per division, $\mu$, for a given simulation and as mentioned earlier, for each cell, we record the number of mutations it accumulates. For each dividing cell, we conduct a Bernoulli trial with the mutation rate and if a mutation occurs, we record its occurrence by updating the mutation count for that cell. We consider two regimes of mutation accumulation-neutral and non-neutral. Neutral mutations do not affect cell fitness, differentiation stage, or any of the differentiation rates, and their accumulation is therefore entirely driven by the overall mutation rate and the frequency of cell division. We assume that non-neutral mutations change the value of cell fitness in its current differentiation stage and simulate this change by replacing $\omega_i$ with mutant fitness, $\omega_i'$. Based on empirical evidence on the distribution of fitness effects in the context of somatic evolution, the mutant fitness is sampled from a lognormal distribution with mean $9 \times 10^{-5}$ and standard deviation $12 \times 10^{-5}$ \citep{eyre-walker2006} and the fitness for subsequent differentiation stages is re-calculated based on Equation \ref{eq1} as $\omega_{i+1}' = \alpha \times \omega_i'$. 

\renewcommand{\arraystretch}{1.2}
\begin{table}[ht]
	\centering
	\begin{tabular}{c c c c}
		\firsthline
		Pace of differentiation & $P$ & $Q$ & $(n_i, n_{i+1})$ \\
		\hline
		Minimum & $0.1$ & $0.9$ & $(172, 28)$ \\
		Equal & $0.5$ & $0.5$ & $(100, 100)$ \\
		Maximum & $0.1$ & $0.1$ & $(28, 172)$ \\
		\lasthline
	\end{tabular}
	\caption[Numerical example of different paces of differentiation]{\label{table1}\textbf{Numerical example of different paces of differentiation}. $P$ and $Q$ are the probabilities of asymmetric and symmetric non-differentiating, division respectively, and the last column gives the average number of progeny cells in stages $i$ and $i+1$ for every 100 cells dividing in stage $i$.}
\end{table}

\subsection{Cell death}
\paragraph{\empty}As with cell division and differentiation, cell death is simulated as a Bernoulli trial, with each cell having a constant probability of 0.0001 of dying in a given iteration of the model. Cells that succeed in the Bernoulli trial are removed from the simulation.

\subsection{Termination of growth}
\paragraph{\empty}We consider two conditions under which cell division is terminated: (1) if $S(t) + n_T(t) \geq M$, then the target size has been reached and the simulation is ended. (2) If  $S(t) + n_T(t)$ is less than the target size $M$, but $S(t)=0$, it implies that the stem cell pool has been depleted before target size could be reached and the simulation is terminated as development is considered unviable (both $S$ and $n_T$ have been defined in a previous section). It is important to note here that since $S$ is the total number of divisible cells across all stages of differentiation, target size $M$ can be reached by increasing $S$ and/or $n_T$, i.e., the number of cells in any differentiation stage from 1 to $T$. Going by condition (1) above, for a given target size, cell compartments could be composed of varying proportions of $S$ and $n_T$ even when development is viable. It also follows from our definitions that $D$ can only be reduced by an increase in the number of terminally-differentiated cells, $n_T$ (Figure \ref{fig1}A). Consequently, $n_T$ plays a key role in determining the rate at which the cell number deficit decreases during growth. Cell compartments of the same size differing in terms of $n_T$ could therefore result in different temporal dynamics of the cell number deficit. Taken together, these two features of the model allow compartment composition to vary independently of compartment size, such that cell compartments composed of different frequencies of stem, transit and differentiated cells can be produced over a range of target sizes. We consider this an essential feature of the model as it enables us to investigate the effects of compartment composition on mutation accumulation as well as how these effects scale with target size.

\subsection{Simulations and parameter exploration}

\paragraph{\empty}The following variables constitute the parameter space over which the behavior of the model will be explored: compartment size, $M$, the cell fitness gradient with differentiation, $\alpha$, pace of differentiation as defined by $P-Q$ combinations in Table \ref{table1}, and the somatic mutation rate, $\mu$.

\paragraph{\empty}This parameter space is arguably vast and several simplifying assumptions are necessary for a tractable study. Due to computational constraints, we explore compartment sizes in the range of $10^3$ to $10^7$. This is a subset of the variation seen across internal organs in humans \citep{nunney1999, tomasetti2015}, and several orders of magnitude below total body sizes across mammalian taxa \citep{Savage2007}. Within this range of compartment size, early simulations showed that mutation rates corresponding to somatic mutation accumulation in human tissues of the order of $10^{-8}$ \citep[and supplementary material thereof]{tomasetti2015} resulted in very low mutation loads with very little scope for multiple mutational hits per cell. We therefore scale up mutation rates in our model to a final range of $10^{-2}$ to $10^{-6}$ per cell per division.

\paragraph{\empty}Empirical data for the fitness gradient $\alpha$ and pace of differentiation are rather limited, and we adopt a general approach for their exploration. As mentioned earlier, $\alpha=1$ leads to no change in cell fitness due to differentiation and can therefore be considered a null case against which other possibilities can be compared. While it is typically assumed that differentiation usually leads to higher division rates (i.e., increasing fitness gradient), there is some empirical evidence that the fitness gradient with differentiation varies more substantially across human tissues \citep{bjerknesAssessmentSymmetryStemcell1985, Morrison2006, lopez-garciaIntestinalStemCell2010}. We therefore explore decreasing, increasing, and flat gradients, corresponding to $\alpha < 1$, $> 1$ and $= 1$, respectively, to cover a range of theoretical possibilities. The $P-Q$ combinations corresponding to the three levels of pace have been discussed in an earlier section (Table \ref{table1}).

\renewcommand{\arraystretch}{1.2}
\begin{table}[ht]
	\centering
	\begin{tabular}{c c c}
		\firsthline
		Trait name & Range & Step size \\ \hline
		\multirow{3}{*}{Pace of differentiation} & Minimum & N.A \\
												& Equal & N.A \\
												& Maximum & N.A \\ \hline	
		Gradient, $\alpha$ & $[0.1, 1.6]$ & $0.1$ \\ \hline
		Somatic mutation rate, $\mu$ & $[10^{-6}, 10^{-2}]$ & Decreasing by an order of magnitude \\ \hline
		Target size, $M$ & $[10^3, 10^7]$ & Increasing by an order of magnitude \\
		\lasthline
	\end{tabular}
	\caption[List of free parameters explored across simulations]{\label{table2}\textbf{List of free parameters explored across simulations}. Detailed explanations of each parameter and the basis for their parametrization are given in the main text.}
\end{table}
 
\paragraph{\empty}Table \ref{table2} summarizes the free parameters explored in this study and for each combination of these parameters, 100 replicate simulations have been run, except for the largest target size for which only 35-62 replicates could be run across levels of pace and somatic mutation rates due to computational constraints. All codes have been written in Python version 3.6 \citep{vanderplas2016} and replicate simulations run on the PARAM Brahma supercomputing facility in IISER Pune using custom SLURM scripts to parallelize replicate runs of a given parameter combination. In the interest of simplicity, we use a fixed value of $T$ for all simulations, but this is undoubtedly likely to be a parameter of interest in further study. 
 
\paragraph{\empty}At the end of each simulation, we record the number of cells $n_i$ in each differentiation stage, $i$. We also record the composition of the compartment across differentiation stages using the weighted average value of differentiation stage, $\overline{\imath}$, given by:

\begin{equation} 
	\label{eq2}
	\overline{\imath} = \frac{\sum_{i=1}^{T} i \cdot n_i}{\sum_{i=1}^{T} n_i}
\end{equation}
	
Cancer progression hinges on sequential accumulation of many mutations in the same cell as well as the clonal outgrowth of those cells. To track this, we record the number of cells $n_m$ that contain $m$ mutations, for $m$ ranging from 1 to 10. Consequently, $m$ represents the mutant clone length, i.e., the number of mutations accumulated in a given cell; a cell that has accumulated 5 mutations over the course of a simulation is said to have reached a clone length of 5. At the population level, the clone length can be measured either in terms of the total clonal growth of mutant cells across clone lengths, or the mutant clone with the highest length. The former is recorded as the weighted average clone length for the cell population $\overline{m}$, given by:

\begin{equation} 
	\label{eq3}
	\overline{m} = \frac{\sum_{m=1}^{10} m \cdot n_m}{\sum_{m=1}^{10} n_m}
\end{equation}

For the latter, we plot $M$ as the maximum value of $m$ in a given run of the model for which $n_m>0$.

\section{Results}
\paragraph{\empty}This study seeks to investigate the variation in patterns of cancer risk allometry across multiple scales of biological organization. Using an individual-based model of the growth of a generalized cell compartment, we first examine the mechanistic basis of how the processes of cell division and differentiation respond to changes in the target size of the compartment, and then relate this to corresponding changes in the patterns of mutation accumulation across target sizes.

\paragraph{\empty}Model results given in the following sections develop this understanding sequentially, starting from the scaling of cell compartment composition with changing target sizes, and then elucidating the effects of cell compartment composition on the scaling of cancer risk. Finally, in the Discussion section, we consider the implications of changing target sizes in the model in terms of changing sizes of cell pools across levels of biological organization from species to organs. Based on current empirical evidence, we identify those processes and parameters that could play a mechanistic role in determining the direction of cancer risk allometry at each level.

\subsection{Pace and gradient limit the effects of target size on final composition}

\begin{figure}[ht]
	\centering
	\includegraphics[width = \columnwidth]{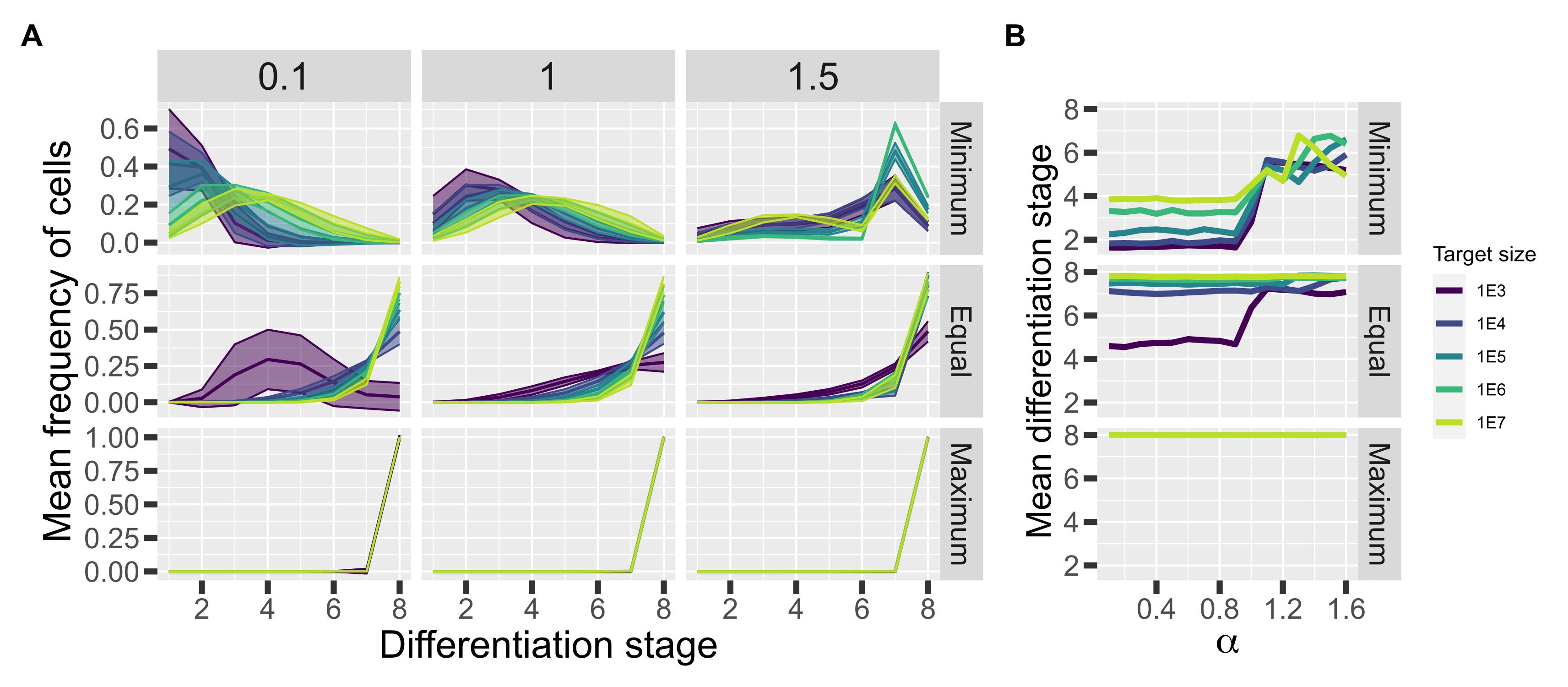}
	\caption[Cell compartment composition as a function of ontogenetic strategy]{\label{fig2}\textbf{Cell compartment composition as a function of ontogenetic strategy}. (A) Relative frequencies of cells against differentiation stages across all target sizes, all three levels of pace and select values of $\alpha$; column headers give the value of $\alpha$ and row headers give the pace level; bold lines are averages over all replicate simulations and ribbons are corresponding SDs (see main text for exact number of replicates). (B) Weighted average differentiation stage against for all three pace levels as given in the row headers, across all target sizes. The calculation of the weighted average differentiation stage is explained in the main text. All values are averaged over replicate runs (see main text for exact number of replicates). Note that the y-axis scale differs across rows.}
\end{figure}

\paragraph{\empty}Figure \ref{fig2}A shows the change in frequencies of cells across differentiation stages for different body sizes, gradients and all three paces of differentiation. From the first two rows, from left to right, we see that as the gradient $\alpha$ increases, the cell populations shift towards greater frequencies of the higher differentiation stages. This bias in the compartment composition is explained readily by how the differentiation gradient affects the choice of cells for division; a steeper gradient, reflected either in very low or very high $\alpha$, leads to selection in favor of either relatively less differentiated or more differentiated cells, respectively. This effect of gradient on the final composition is seen even more clearly in Figure \ref{fig2}B, which shows the mean differentiation stage weighted by the corresponding cell frequencies (Equation \ref{eq2}). Here, we see that as gradient increases, the weighted mean increases, following the shift in cell frequencies from lower to higher differentiated stages, with a clear transition at $\alpha=1$. Moreover, comparing the cell frequency curves for a given target size across different levels of differentiation pace (for a given column across the rows in Figure \ref{fig2}A and B), it is clear that greater pace also increases the general push towards differentiation. The most drastic demonstration of this can be seen in the bottom row of Figure \ref{fig2}, where maximum pace swamps over all other parameters to produce organisms composed almost exclusively of terminally-differentiated cells. A significant proportion of such organisms are not developmentally viable and fail to reach the target size as they exhaust the divisible cell pool before maturity.

\paragraph{\empty}Figure \ref{fig2}B also shows that the target size has a clear effect on the final composition-larger target sizes produce a greater push towards differentiation and a greater proportion of differentiated cells relative to the stem cells. However, the degree to which this push is realized is dependent on both $\alpha$ and the pace of differentiation. Figure \ref{fig2}B visualizes this dependence very clearly, in which maximum pace once again serves as an extreme example of target size having no practical effect on final composition. It also shows the limiting effect of pace on the extent of differentiation with target size as larger target sizes progress to much more differentiation under equal pace than minimum pace. Likewise, the transition in body composition around $\alpha=1$ is clearly conserved under both equal and minimum pace, albeit more clearly for smaller target sizes. We therefore surmise that while higher target size does lead to a generally higher propensity for differentiation, ontogenetic properties like pace and $\alpha$ have the potential to act as tuning knobs on how final cellular composition responds to changes in target size.

\subsection{Deficit dynamics explain how gradient and pace give rise to different cellular compositions}

\begin{figure}[!ht]
	\centering
	\includegraphics[width = \columnwidth]{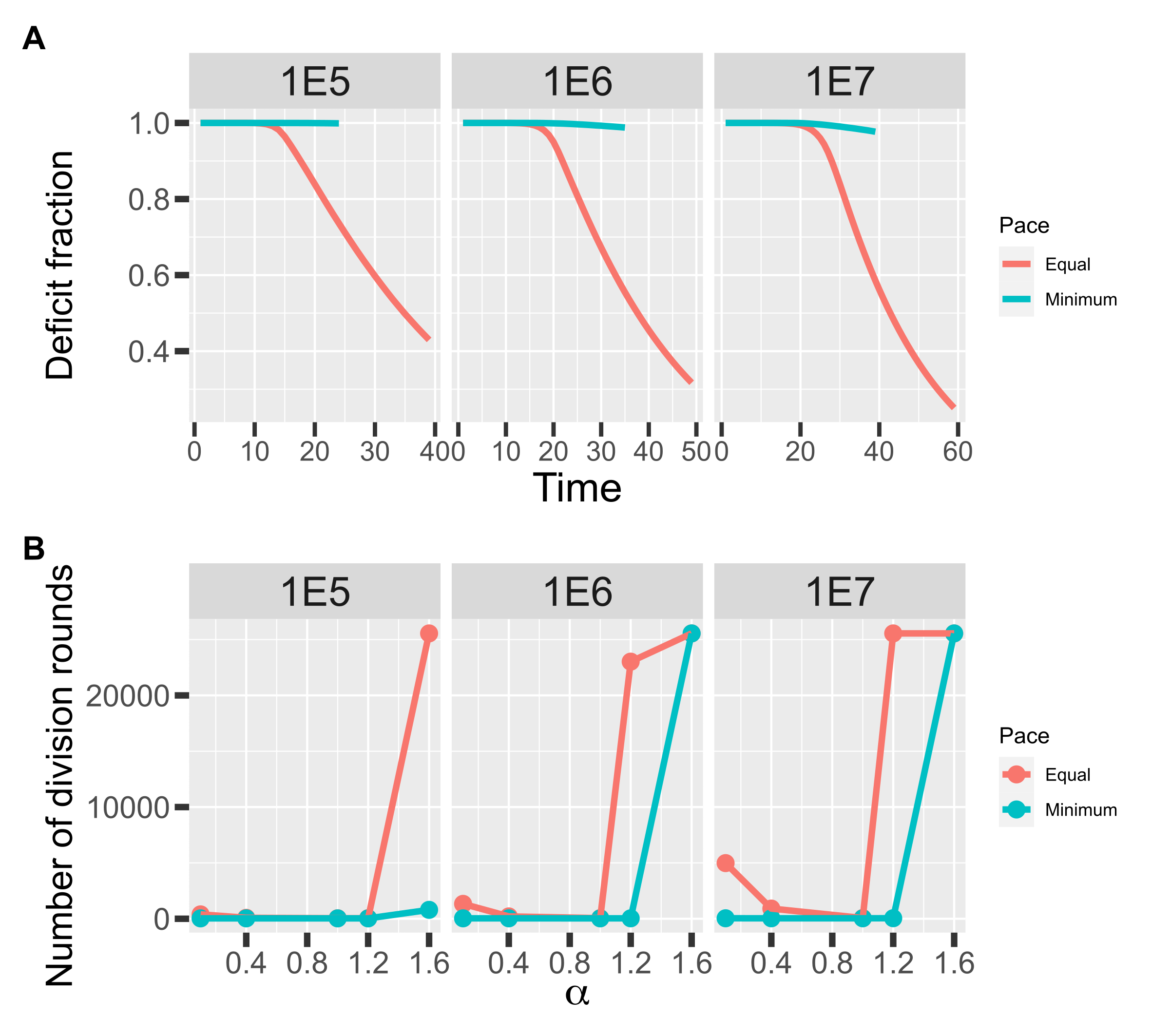}
	\caption[Conditions for tapering deficit dynamics]{\label{fig3}\textbf{Conditions for tapering deficit dynamics}. (A) Deficit dynamics against time for two levels of pace and three target sizes, showing equal pace has a decreasing deficit in time whereas minimum pace maintains a higher deficit for a longer time. (B) Number of rounds of division simulated over the course of development in the model; large values of $\alpha$ and higher pace are clearly associated with more rounds of division, which in combination with a small deficit, implies a progressively smaller dividing pool of cells. For larger target sizes, number of rounds of division increases for very low values of $\alpha$ as well. All lines are averages over 9-11 replicate simulations.}
\end{figure}

\paragraph{\empty}The previous section demonstrated that $\alpha$ and pace influence the extent of differentiation. As the deficit in our model is defined as the difference between the target size and number of terminally-differentiated cells, this could also affect how the deficit changes with time. We examine this here by considering how $\alpha$ and pace affect the temporal dynamics of the cell number deficit and cell division in the model. Figure \ref{fig3}A shows that the cell number deficit declines relatively faster under equal pace of differentiation than under minimum pace; we omit considering maximum pace level here as differentiation progresses too quickly. This decline in cell number deficit corresponds to the trends in the number of rounds of cell division required to reach maturity-as shown in Figure \ref{fig3}B, equal pace simulations consistently undergo as many or more rounds of division than minimum pace. Likewise, we see that extreme values of $\alpha$, both very low and very high, entail more rounds of cell division than intermediate values of $\alpha$. $\alpha$ and pace therefore influence how division happens in the model, and specifically, how many divisions are required to meet the target size. Understanding these effects of pace and $\alpha$ requires a closer look at the workings of the model.

\paragraph{\empty}Our selection algorithm begins by ranking cells by fitness, so we start with a sequence of cells arranged in decreasing order of fitness. Selection could be visualized as a ``window'' placed on this ordered sequence such that cells within the window are selected for further division. The deficit corresponds to the size of this window, and a larger window leads to more cells being selected. Since $\alpha$ represents the scaling of cell fitness with differentiation (Figure \ref{fig1}B), it determines which differentiation stage is highest in fitness. This, in turn, determines the differentiation stage of most of the cells at the top of the fitness hierarchy which are included in the selection window for subsequent division. Pace determines the number of further differentiated cells produced per division event, and lower pace leads to lesser overall differentiation. Consequently, when differentiation pace is low, cells in the selection window are more likely to be from a narrow range of differentiation stages if not a single differentiation stage and by corollary, higher pace leads to a more diverse group of differentiated and less differentiated cells within the selection window. Based on this logic, it can be seen that minimum pace and small values of $\alpha$ together constitute a strategy for producing preferentially less differentiated cells, whereas higher pace and higher $\alpha$ both increase the push towards differentiation. We refer to these as ``low differentiation'' and ``high differentiation'' strategies respectively, based on the net progression of differentiation in each case. As the cell number deficit is only decreased by the accumulation of terminally-differentiated cells ($n_T$), parameter combinations corresponding to high differentiation strategies are also more likely to reduce the deficit in time, whereas those that restrict differentiation (low differentiation strategies) would maintain a large deficit for much of the simulated time. Furthermore, as deficit remains high for low differentiation strategies, growth happens in saltatory leaps forward and the target size is eventually reached or even exceeded with most cells being from the lower differentiation stages. On the other hand, high differentiation strategies lead to progressive reductions in deficit with time and the target size is reached almost asymptotically, with fewer and fewer cells selected in each iteration, and the target size approached in smaller steps.

\paragraph{\empty} The parameter combination of equal pace and very low values of $\alpha$ does not strictly fall in this spectrum of high and low differentiation strategies as the two parameters act somewhat in opposition to each other. However, very low values of $\alpha$ favor less differentiated cells for division, whereas equal pace leads to a higher frequency of differentiation. The selection window is placed around the lowest differentiation stages, which are progressively depleted through differentiation. The size of the divisible cell pool therefore reduces progressively, and as with the high differentiation strategies, the target size is consequently approached in smaller steps as fewer cells are available for division.

\paragraph{\empty}Taken together, this shows that pace and $\alpha$, by changing the way in which cells are selected for division and the extent to which cell division contributes to differentiation, can affect the cellular composition of the compartment at maturity. It is also worth noting here that this effect is more prominent for larger target sizes, where the model requires more cell divisions to reach the target size, with the deficit persisting for longer.

\subsection{A reducing deficit increases accumulation of neutral mutations}

\begin{figure}[!ht]
	\centering
	\includegraphics[width = \columnwidth]{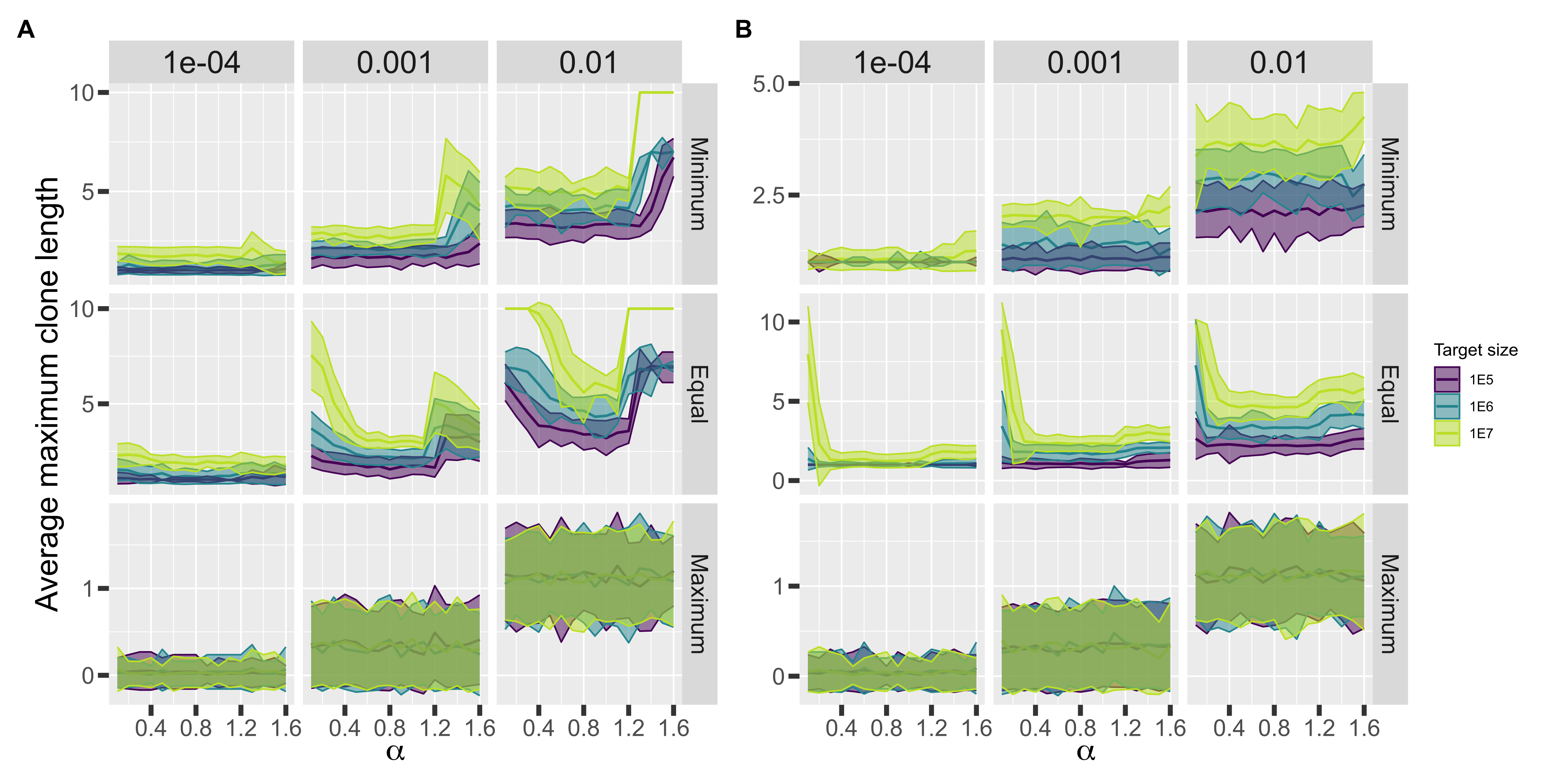}
	\caption[Maximum clone length corresponds to conditions of tapering deficit dynamics]{\label{fig4}\textbf{Maximum clone length corresponds to conditions of tapering deficit dynamics}. Maximum clonal length as defined in the main text against under (A) neutral and (B) non-neutral mutation accumulation. Columns are somatic mutation rates, and rows are pace levels. Plotted values are averages over all replicate simulations (see main text for exact number of replicates). Note that the y-axis ranges change from (A) to (B), and across rows.}
\end{figure}

\paragraph{\empty}Given that mutation in our model is intrinsically linked to cell division, changes in cell division patterns could lead to altered patterns of mutation accumulation as well. This is highly relevant for cancer risk, as much of cancer origin and progression is thought to occur through sequential accumulation of multiple mutations within a single cell or a small group of cells \citep{Frank2007, calabrese2010, nunney2015}. Figure \ref{fig4} shows the average maximum clone length as defined in Equation \ref{eq3} for three values of the somatic mutation rate across columns. Here, we consider both neutral mutations that do not affect cell fitness (Figure \ref{fig4}A), as well as non-neutral mutations that influence the fitness (Figure \ref{fig4}B). Sequential accumulation of neutral mutations is affected by the number of rounds of division such that, where more rounds of division are required to reach maturity, maximum clone lengths also tend to be higher on average (\textit{cf} Figure \ref{fig3}B and Figure \ref{fig4}A). Strategies that entail more rounds of division therefore also lead to more sequential mutation accumulation. Furthermore, we see that increasing target size likewise increases the maximum clone length regardless of the differentiation strategy. This is an important observation as it represents the direct effect of the scaling of cell division with target size on the baseline risk of mutation, independent of somatic evolution driven by the fitness of mutant clones.

\paragraph{\empty}We return to our earlier explanation of cell division in the model to understand these trends in the accumulation of neutral mutations. As identified earlier, high differentiation strategies, as well as equal pace with very low $\alpha$, both entail more rounds of division due to a reducing deficit and a slow approach toward the target size (Figure \ref{fig3}). As the deficit declines, it also leads the selection window to shrink, so that as the target size is approached, a small set of cells with the highest fitness are selected repeatedly and undergo several rounds of division. The higher frequency of division of a small cell pool therefore creates more opportunities for mutations over a short period of time, and particularly for multiple mutational hits in the same set of cells. On the other hand, low differentiation strategies are not associated with a declining deficit (Figure \ref{fig3}) and the selection window remains large throughout the growth period. The pool of actively dividing cells is large enough that the same cells are less likely to be selected repeatedly for division, and a large dividing cell pool further results in fewer rounds of cell division in general to reach the target size. Both these effects make the sequential accumulation of multiple mutations in the same cell less likely. The maximum clone length therefore closely follows those growth strategies that produce a declining deficit.

\paragraph{\empty}Two more trends are important to point out in the context of mutation accumulation. First, increasing the target size and the somatic mutation rate both have qualitatively similar effects on the maximum clone length, as clone length generally increases in either case (Figure \ref{fig4}A). Second, our data also indicate that a higher clone length does not automatically imply higher mutant frequencies in the cell pool (Figure \ref{fig6}). This implies that mutation accumulation and mutant clone growth are independent processes, and while the differentiation strategy affects the risk of mutation accumulation, the actual expansion of potentially oncogenic mutations probably occurs over much longer timescales than what we have considered in our model. This is consistent with the general understanding that cancer progression involves both the accumulation and subsequent growth of oncogenic mutations \citep{Aktipis2013}.

\subsection{Non-neutral mutation accumulation reveals a balance between differentiation and mutation}
\paragraph{\empty}The understanding of neutral mutation dynamics developed above can now be extended to elucidate the accumulation of non-neutral mutations. The distribution of fitness effects of mutations in our model generates largely deleterious mutations, such that large-effect beneficial mutations are relatively rare. Comparing Figures \ref{fig4}A and B, we see that in most parts of the parameter space, the trends in maximum clone length are qualitatively similar between neutral and non-neutral mutation accumulation, but the latter shows a markedly lower frequency of mutations overall with shorter clone lengths. This is consistent with the fact that deleterious mutations, which are by far more common in the model, are filtered out of the cell population by selection, which leads to fewer mutations on the whole.

\paragraph{\empty}However, low $\alpha$ under equal pace presents a unique region in the parameter space in terms of non-neutral mutation accumulation, as it shows particularly large maximum clone lengths (Figure \ref{fig4}B, middle row). As it has been pointed out earlier, this parameter combination leads to qualitatively similar dynamics of the cell number to high differentiation strategies, in that it leads to a progressively shrinking pool of actively-dividing cells that undergo many more rounds of division (Figure \ref{fig3}B). Based on this similarity, patterns in non-neutral mutation accumulation under low $\alpha$ and equal pace are expected to be similar to those in high differentiation strategies. However, the data indicate that non-neutral mutation accumulation under low $\alpha$ and equal pace run contrary to this expectation, showing much larger maximum clone lengths than high differentiation strategies (\textit{cf} Figure \ref{fig4}A and B, middle rows). This discrepancy is intriguing as two growth strategies with seemingly similar dynamics of the cell number deficit show divergent trends in the accumulation of mutations. In order to understand the cause for such a disparity, it is necessary to examine the relative frequencies of non-mutant and mutant cells for low $\alpha$ with equal pace, which clearly shows a much higher frequency of mutant cells with multiple mutations (Figure \ref{fig6}, middle rows). This arises as a consequence of the fact that under low $\alpha$ and equal pace, high-fitness mutations can potentially compensate for the reduction in cell fitness due to differentiation. Therefore, the cell number deficit can be reduced by higher frequencies of differentiated cells with a high extent of mutation. Since very low values of $\alpha$ lead to a drastic reduction in cell fitness with differentiation, multiple mutations are more likely to increase cell fitness sufficiently for them to be included in the selection window. This therefore indicates that under non-neutral mutation accumulation, the cell number deficit is reduced by different sets of cells under low $\alpha$ and equal pace than under high differentiation strategies. The former fills the deficit through the division of differentiated and multiply-mutant cells, while the latter fills the deficit through the division of undifferentiated, non-mutant cells. Low $\alpha$ and equal pace thus allows for antagonistic action between differentiation and mutation through their opposing effects on cell fitness. This antagonism allows the low $\alpha$ and equal pace strategy to produce divergent patterns in the accumulation of non-neutral mutations despite the dynamics of cell number deficit being qualitatively similar to that of high differentiation strategies.

\subsection{Viability corresponds to depletion of the dividing cell pool}

\begin{figure}[!ht]
	\centering
	\includegraphics[width = \columnwidth]{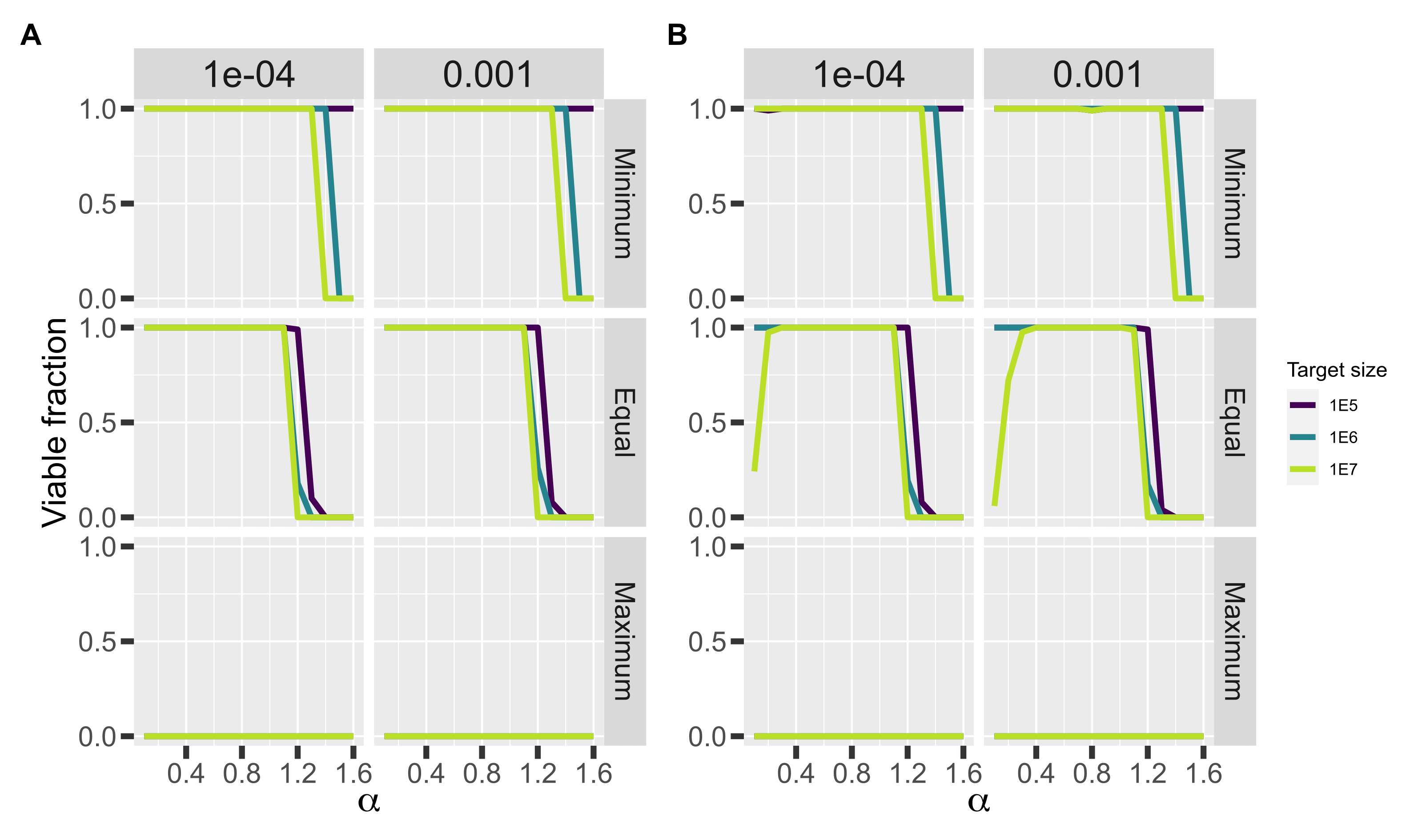}
	\caption[developmental viability is related to mutation accumulation]{\label{fig5}\textbf{Developmental viability is related to mutation accumulation}. Fraction of viable model runs in which the target size was attained without stem cell depletion under (A) neutral and (B) non-neutral mutation accumulation. $\alpha$ is on the x-axis here, columns are somatic mutation rates and rows are pace levels. Plotted values are averages over all replicate simulations (see main text for exact number of replicates).}
\end{figure}

\paragraph{\empty}For the purposes of this study, we consider a given instance of a simulation to be viable if the total cell number reaches the target size within a margin of 10\% lower or higher than the actual target size. From Figure \ref{fig5}, it is clear that under both neutral and non-neutral mutation accumulation, viability reduces consistently in regions of the parameter space corresponding to high differentiation strategies, namely high $\alpha$ and/or high pace. Figure \ref{fig2} clearly shows that high differentiation strategies also lead to compartments with very low frequencies of the lower differentiation strategies. Particularly for the largest target sizes, this leads to an insufficient number of divisible cells for the target size to be reached.

\begin{figure}[!ht]
	\centering
	\includegraphics[keepaspectratio=true, height=0.35\textwidth]{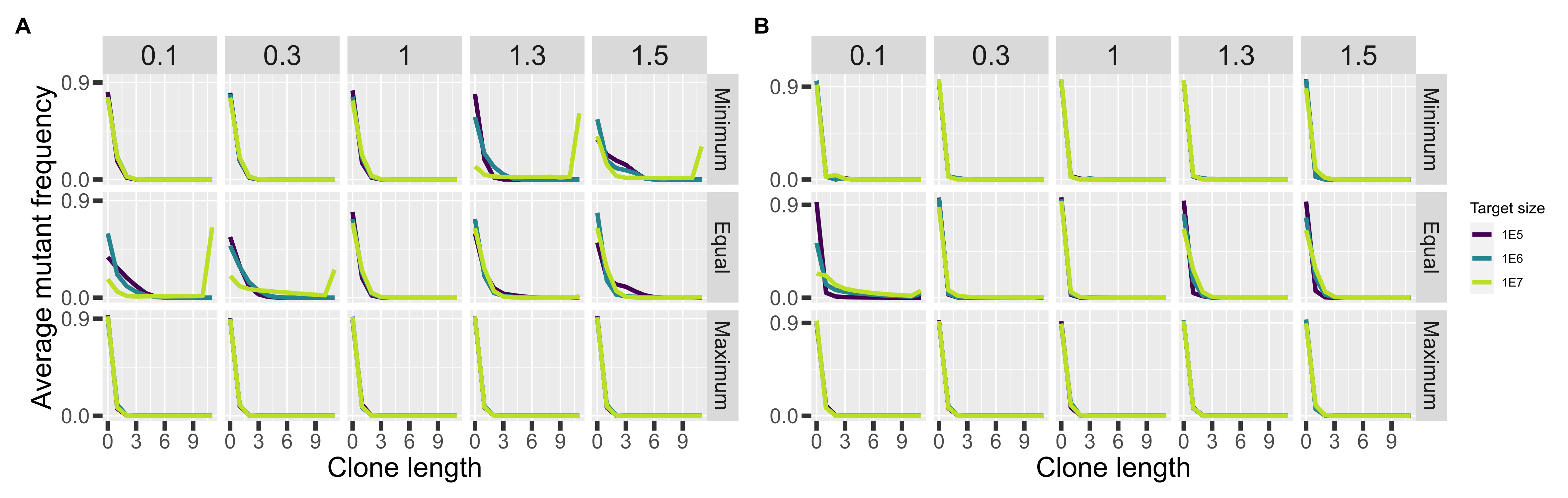}
	\caption[Mutant clone growth across model parameters]{\label{fig6}\textbf{Mutant clone growth across model parameters}. Relative cell frequency against clone length, which is the number of mutations in a given cell under (A) neutral and (B) non-neutral mutation accumulation. All plots correspond to mutation rate, $\mu=0.01$ per cell per division. As with previous figures, pace of differentiation varies across rows and  across columns. Replicate numbers are the same as in Figure \ref{fig2}.}
\end{figure}

\paragraph{\empty}Furthermore, viability also drops under non-neutral mutation accumulation for low values of $\alpha$, which corresponds to the unusually high clone lengths pointed out earlier (\textit{cf} Figures \ref{fig5}B and \ref{fig4}B). It follows from the previous explanation that this increase in clone length excludes differentiated non-mutant cells in favor of less differentiated non-mutants and high-fitness mutants. This effectively shrinks the pool of cells available for division over the entire course of the simulation, making it more likely that divisible cells are exhausted before the target size is attained.

\section{Discussion}

\paragraph{\empty}We have shown here that while cellular composition is a composite function of $\alpha$, the pace of differentiation, and the target size, patterns in sequential mutation accumulation across these parameters can actually be understood in terms of the temporal dynamics of the cell number deficit over the course of ontogeny. More specifically, we identify so-called high differentiation strategies that are consistently associated with greater propensity for differentiation over the course of cell division (Figure \ref{fig2}) and a small cell number deficit over an extended period of time (Figure \ref{fig3}). High-differentiation strategies consequently involve the repeated selection of a small pool of high-fitness cells that undergo a disproportionately higher number of rounds of cell division. We also show that this repeated division of the same set of cells can be related directly to the sequential accumulation of several mutations in the same cell (Figure \ref{fig4}). It bears repeating here that we take mutation accumulation as a direct proxy for cancer risk, in accordance with the current understanding of the centrality of mutation in the origin and progression of cancer. The inferences from our data regarding mutation accumulation applies to this general notion of cancer risk, such that greater mutation accumulation in some region of the parameter space indicates a higher risk of cancer for the corresponding combination of parameter values. So far, we have been discussing our findings purely in terms of how the various model parameters function to produce the patterns of cell division and mutation, and their scaling with target size. This general mechanistic understanding can now be extended to the cancer risk trends across biological scales mentioned earlier, and at each level of biological organization, we can use the model findings to speculate about mechanisms to explain the emergence of corresponding trends in cancer risk. More specifically, at each level of biological organization, we recast the known trends in cancer risk at that level in terms of the plausible action of our model parameters as applicable, in order to identify model-based mechanisms that are consistent with the cancer risk trend in question. This leads us to predictions regarding which parameters are likely to be of mechanistic importance across biological scales and therefore merit more focused investigation.

\paragraph{\empty}We begin at the highest organizational level, with the trends in cancer risk across species of varying body sizes. Within this biological level, cancer risk trends have been studied most extensively among mammalian species, and we therefore place our results in the context of cancer risk scaling among mammals \citep{Abegglen2015}. As mentioned earlier, empirical data regarding how cellular composition changes with body size across mammals are rather sparse. While it is known that some cell types become larger with increasing body size among birds \citep{Kozlowski2010}, and to some extent, mammals \citep{Savage2007}, it is unclear if the actual number of stem cells and/or their proportion relative to differentiated cells change systematically with body size. Results from EK20 suggest that an increase in the size of the stem cell compartment could be possible in larger species \citep{Erten2020}. In the context of our model, sustaining a larger stem cell compartment requires a reduced progression of differentiation, which would be facilitated most by small values of $\alpha$. Moreover, for any given value of $\alpha$, higher pace of differentiation exhausts the stem cell compartment entirely, leaving very few cells in the lowest two differentiation stages (Figure \ref{fig2}). This indicates a potential upper limit for the pace of differentiation as body size increases across species, such that larger organisms would need to limit the rate of ontogenetic asymmetric divisions to prevent depletion of the stem cell compartment. On the other hand, we also find that $\alpha << 1$, while supporting a large stem cell compartment, also leads to significantly longer clonal lineages and higher mutant frequencies (Figures \ref{fig4} and \ref{fig6}). Sequential mutation accumulation could therefore be the basis for a trade-off between $\alpha$ and the pace of differentiation in determining the scaling of cellular composition with body size. Since we consider sequential mutation accumulation a proxy for cancer, we speculate that lower pace of differentiation and intermediate values of $\alpha$ are likely to be consistent with increasing body sizes across species without a proportionate increase in cancer risk. The simple case of non-neutral mutation accumulation that we tested (Figure \ref{fig4}B) does not seem to contradict this expectation as mutant lineages become dramatically longer with body size only for equal pace and small $\alpha$. 

\paragraph{\empty}One step lower on the organizational ladder, the scaling of cancer risk with body size between individuals of the same species can also be explained based on our model results. Since $\alpha$ and the pace of differentiation are both ontogenetic variables, it is plausible to assume that for a given species, their values remain constant from one individual to the next. Based on this assumption, we see that for a given value of $\alpha$ and differentiation pace, the extent of mutation accumulation, both in terms of clonal length and to a limited degree, relative mutant frequencies, does in fact increase with increasing target size (Figures \ref{fig4} and \ref{fig6}). Within this regime, mutation accumulation seems entirely limited by cell number and mutation rate, and increasing either one of the two increases mutation risk. The correspondence of these trends from our data with empirical observations of cancer risk is straightforward, as cancer risk is known scale with body size within a given species \citep{albanes_adult_1988, green2011, nunney2018, rafalko2023}. Our model data are therefore consistent with the possibility that ontogenetic parameters could be conserved between conspecifics. On the other hand, our data also suggest that larger individuals of the same species should have a greater flux towards differentiation and therefore end up with a higher proportion of more differentiated cells than smaller conspecifics. To our knowledge, empirical data have not been collected at such resolution within any species, which precludes further comments on the question.

\paragraph{\empty}Organ-level cancer risk arguably involves variation in almost all our model parameters, including the somatic mutation rate. Research over the past few years has shown that while much of the variation in cancer risk between organs could be ascribed to the number of lifetime stem cell divisions in each tissue \citep[lscd, \textit{sensu}][\empty]{tomasetti2015, Tomasetti2017}, tissue-specific effects can shift the cancer risk-lscd correlation along the x-axis and/or change its slope \citep{Noble2015}. As suggested by Noble and Hochberg \citep{Hochberg2017}, such tissue-specific effects could stem from differential exposure to environmental risk factors leading to tissue-specific somatic mutation rates. But apart from these external influences, different tissues have different target sizes, and to some extent, organizational features like the architecture of the stem cell compartment are also known to vary across human tissues \citep{bjerknesAssessmentSymmetryStemcell1985, Morrison2006, lopez-garciaIntestinalStemCell2010}. Our model data reflect part of such differences, such that different tissue organizations across organs could be consistent with different values for $\alpha$ and differentiation pace in the model. On the other hand, we do not vary the number of transit stages, $T$ between the basal stem cell and differentiated cells, even though $T$ is thought to differ across tissues \citep{demeter2022}. This is a key limitation for mutation accumulation in particular as the tapering dynamics of the cell number deficit shown in Figure \ref{fig3} frequently occur within the highest few differentiation stages and the addition of more stages could postpone the tapering of the deficit sufficiently to allow the target size to be reached with fewer mutations. EK20 also find that higher $T$ is also a component of the optimal ontogenetic strategy that supports evolutionary increases in target size \cite{Erten2020}. Future theoretical work would therefore need to be targeted at studying the effects of changing $T$ on ontogenetic dynamics across organs to elucidate any factors that determine cancer risk in a tissue-specific manner.

\paragraph{\empty}Finally, the tapering of deficit towards the end of ontogenetic growth suggests an interesting possibility for the accumulation of mutations over the course of post-maturity organismal life. It has been shown that mutations accumulate nearly linearly with age in most tissues due to cell divisions as part of baseline turnover \citep{blokzijl2016}. While we do not explicitly simulate aging and lifetime somatic maintenance in our model, long-term cell turnover could be comparable to a constant small deficit over a long period with no change in any of the other parameters. Under such circumstances, unless the mutation rate itself changes, mutation accumulation would be expected to proceed more or less monotonically with further lifetime cell turnover as division continues to be concentrated within a small pool of dividing cells. A recent study has shown that somatic mutation rates across mammalian species scale negatively with lifespan but not body size \citep{caganSomaticMutationRates2022}, which suggests that the prolonged cell turnover likely has a stronger role to play in determining mutation accumulation than a larger cell pool. \textit{A priori}, this is consistent with the observed effects of a tapering cell number deficit on mutation accumulation in the model across target sizes. On the other hand, variables like the pace of differentiation could also change with age due to senescence-related changes in the stem cell micro-environment \citep{Liggett2017a}, indicating that this is a rather simplistic view that is only a starting point for further investigation.

\paragraph{\empty}At least two caveats are worth pointing out. First, the current study has evaluated cancer risk scaling with ontogenetic strategies purely in terms of mutation occurrence. As has been mentioned earlier, mutant clone length only reflects the occurrence of mutations while mutant clone growth not always following immediately. This difference is highly relevant for the temporal dynamics of cancer progression and detection, and should therefore be explored in future work. Second, the assumption that the deficit $D$ is reduced only by $n_T$ is a modelling choice that is not rooted in any rigorous biological observations. Nevertheless, it is not unreasonable to suppose that all organisms must maintain some minimum number of terminally-differentiated cells in order to perform various physiological activities. Functional maturity can therefore potentially by reflected only in $n_T$. This is a speculative argument, and to truly address this objection, more data are needed that delineate the dynamics of metazoan development in greater detail. In the meantime, the current model still provides a reasonable theoretical framework based on a general understanding of how mammals develop to augment the available data on cancer risk trends with mechanistic insights and testable hypotheses.

\bibliographystyle{plainnat}
\bibliography{thesis-references}

\newpage
\section*{Funding support}
BV was supported by IISER Pune. The authors declare no conflicts of interest.

\section*{Acknowledgements}
The authors are grateful to Prof Sutirth Dey and members of the Population Biology Lab (PBL) at IISER Pune for their support and feedback. BV also acknowledges critical inputs from Prof TNC Vidya (JNCASR Bengaluru).

\section*{Data availability}
The authors undertake to make all relevant codes and data available publicly upon accceptance of the manuscript.
\end{document}